\documentclass[prb,superscriptaddress, twocolumn]{revtex4-1}
\usepackage{mathtools,multirow,amssymb}
\usepackage[T1]{fontenc}
\usepackage{gensymb}
\usepackage{graphicx}

\usepackage{graphicx,rotating,booktabs}
\usepackage[verbose]{placeins}
\usepackage{blindtext}

\begin{document}

\title{Stability and elasticity of metastable solid solutions and superlattices in the MoN--TaN system: a first-principles study}

\author{Nikola Koutn\'a}
\affiliation{Institute of Materials Science and Technology, TU Wien, Getreidemarkt 9, A-1060 Vienna, Austria}
\affiliation{Department of Condensed Matter Physics, Faculty of Science, Masaryk University, Kotl{\' a}{\v r}sk{\' a} 2, CZ-611 37 Brno, Czech Republic}

\author{David Holec}
\affiliation{Department of Physical Metallurgy and Materials Testing, Montanuniversit\"{a}t Leoben, Franz-Josef-Strasse 18, Leoben A-8700, Austria}

\author{Martin Fri\'{a}k}
\affiliation{Institute of Physics of Materials, Academy of Sciences of the Czech Republic, v.v.i., \v{Z}i\v{z}kova 22, CZ-616 62 Brno, Czech Republic}
\affiliation{Central European Institute of Technology, CEITEC MU, Masaryk University, Kamenice 5, CZ-625 00 Brno, Czech Republic}
\affiliation{Department of Condensed Matter Physics, Faculty of Science, Masaryk University, Kotl{\' a}{\v r}sk{\' a} 2, CZ-611 37 Brno, Czech Republic}

\author{Paul H. Mayrhofer}
\affiliation{Institute of Materials Science and Technology, TU Wien, Getreidemarkt 9, A-1060 Vienna, Austria}

\author{Mojm\'ir \v{S}ob}
\affiliation{Central European Institute of Technology, CEITEC MU, Masaryk University, Kamenice 5, CZ-625 00 Brno, Czech Republic}
\affiliation{Institute of Physics of Materials, Academy of Sciences of the Czech Republic, v.v.i., \v{Z}i\v{z}kova 22, CZ-616 62 Brno, Czech Republic}
\affiliation{Department of Chemistry, Faculty of Science, Masaryk University, Kotl{\' a}{\v r}sk{\' a} 2, CZ-611 37 Brno, Czech Republic}

\begin{abstract}
Employing {\it{ab initio}} calculations, we discuss chemical, mechanical, and dynamical stability of MoN--TaN solid solutions together with cubic-like MoN/TaN superlattices, as another materials design concept. 
Hexagonal-type structures based on low-energy modifications of MoN and TaN are the most stable ones over the whole composition range. 
Despite being metastable, disordered cubic polymorphs are energetically significantly preferred over their ordered counterparts. 
An in-depth analysis of atomic environments in terms of bond lengths and angles reveals that the chemical disorder results in (partially) broken symmetry, i.e., the disordered cubic structure relaxes towards a hexagonal NiAs-type phase, the ground state of MoN. 
Surprisingly, also the superlattice architecture is clearly favored over the ordered cubic solid solution. We show that the bi-axial coherency stresses in superlattices break the cubic symmetry beyond simple tetragonal distortions and lead to a new tetragonal $\zeta$-phase (space group P4/nmm), which exhibits a more negative formation energy than the symmetry-stabilized cubic structures of MoN and TaN. 
Unlike cubic TaN, the $\zeta\text{-TaN}$ is elastically and vibrationally stable, while $\zeta$-MoN is stabilized only by the superlattice structure. 
To map compositional trends in elasticity, we establish mechanical stability of various Mo$_{1-x}$Ta$_x$N systems and find the closest high-symmetry approximants of the corresponding elastic tensors. 
According to the estimated polycrystalline moduli, the hexagonal polymorphs are predicted to be extremely hard, however, less ductile than the cubic phases and superlattices. 
The trends in stability based on energetics and elasticity are corroborated by density of electronic states.
\end{abstract}

\maketitle

\section{Introduction}
Transition metal nitrides (TMNs) represent a prominent class of materials possessing numerous outstanding physical properties, such as excellent chemical and thermal stability, incompressibility and strength, high melting point, good thermal and electric conductivity or superconductivity \cite{Liu2012structure, Liu1989prediction, He2004hardness, PhysRevB.79.024111}. 
In order to enhance the performance of these materials, considerable efforts have been devoted to investigate the possibility of fine tuning the mechanical and/or electrical properties by designing ternary or multinary TMN systems \cite{zhou2016structural, klimashin2015composition,djemia2013structural,holec2012trends,tasnadi2012ab,friak2015synergy}.

The addition of nitrogen atoms into the high-density electronic gas of transition metals together with the covalent bonding to nitrogen atoms leads to extraordinary hardness \cite{Li_Jing_Guo, Calka, Friedrich}. 
For example, the hardness of MoN ranges from 28 to 34\,GPa \cite{Jauberteau2015Molybdenum}, while for TaN it
ranges from 30 to 32\,GPa \cite{Shin1999Growth,Kieffer1974Neue}. 
According to Teter's empirical correlation \cite{teter1998computational}, hardness scales with shear modulus.
Later, \citet{chen2011modeling} proved that hardness also correlates with bulk modulus.
The {\it{ab initio}} calculated bulk moduli are 392\,GPa \cite{Lowther2004Lattice} and 348\,GPa \cite{Chang2012Structure} for the hexagonal ground states of MoN and TaN, i.e., NiAs-type MoN (NiAs prototype, P6$_3$/mmc, $\#194$, conventionally termed as $\delta\text{-MoN}$) and TaN-type TaN (TaN prototype, P$\overline{6}$2m, $\#189$, conventionally termed as $\pi\text{-TaN}$), respectively. 
These values are comparable with the 370\;GPa of the cubic boron nitride\cite{haines2001synthesis} or even with the 443\,GPa of diamond \cite{mcskimin1972elastic}, the hardest material to date. 

Despite being metastable, the cubic modifications of MoN and TaN (Fm$\overline{3}$m, $\#225$), referred to as rocksalt structure (rs), have been synthesized using non-equilibrium growth techniques such as reactive magnetron sputtering in high nitrogen partial pressure atmosphere, nitrogen ion implantation, or low energy ion assisted deposition \cite{Jauberteau2015Molybdenum, Ihara, Mndl2004, Savvides, mashimo1997b1,ensinger1995low,klimashin2016impact,chen1999low}.
Importantly, some properties of the metastable cubic variants are comparable with, or even superior to those of the ground state phases, e.g., cubic TaN prepared by shock and static compression was shown to have very good high-temperature stability comparable to that of the hexagonal WC-type phase \cite{mashimo1997b1}. Other experimental studies \cite{Savvides,inumaru2006structural} suggest that cubic molybdenum nitride is promising as a superconductor. The {\it{ab initio}} predicted transition temperature for the rs-MoN is 29\,K \cite{Papaconstantopoulos}, which is the highest of all refractory carbides and nitrides \cite{Jauberteau2015Molybdenum}.

Although considerable effort has been devoted to $\text{Mo--N}$ and Ta--N, the quasi-binary MoN--TaN system has been rarely studied \cite{bouamama2015first,junhua2014microstructures}. 
Restricting only to cubic system, \citet{bouamama2015first} performed DFT calculations on both (fully) ordered and disordered Mo$_{1-x}$Ta$_x$N. 
The virtual crystal approximation (VCA) used to model the disordered phase in their study, however, is the simplest approach for dealing with alloying effects, since it neglects any possible short range order. 
Therefore, it is desirable to employ more sophisticated approaches closer to reality, e.g., the special qua\-si\-ran\-dom structure (SQS) method \cite{Wei1990Electronic}.

In the present study, we systematically investigate phase stability and elastic properties of Mo$_{1-x}$Ta$_x$N alloys adopting cubic NaCl-type (Fm$\overline{3}$m, rocksalt) or hexagonal NiAs-type (P6$_3$/mmc, ground state of MoN), WC-type (P$\overline{6}$m2, low-energy phase of MoN), and TaN-type (P$\overline{6}$2m , ground state of TaN) structures. 
Considering both (partially) ordered and the SQS-generated disordered structures alongside with MoN/TaN superlattices as another materials design concept, a careful structural analysis of the fully relaxed systems is carried out and discussed in the context of their chemical stability and elastic properties.

\section{Calculation details} \label{Sec: Calculation details}
The Vienna Ab-initio Simulation Package (VASP) \cite{Kresse1996Efficient, Kresse1999From} was used to perform the DFT calculations, employing the projector augmented plane wave (PAW) pseudopotentials under the generalized gradient approximation (GGA) \cite{Kohn1965Self} with a Perdew-Burke-Ernzerhof (PBE) exchange and correlation functional \cite{PhysRevLett.77.3865}. 
The plane-wave cut\-off energy was always set to 700\,eV and the $k$-vector sampling of the Brillouin zone was checked to provide a total energy accuracy of about $10^{-3}\,\mathrm{eV/at}$. 

\begin{figure}[h!]
	\centering
    \includegraphics[width=6cm]{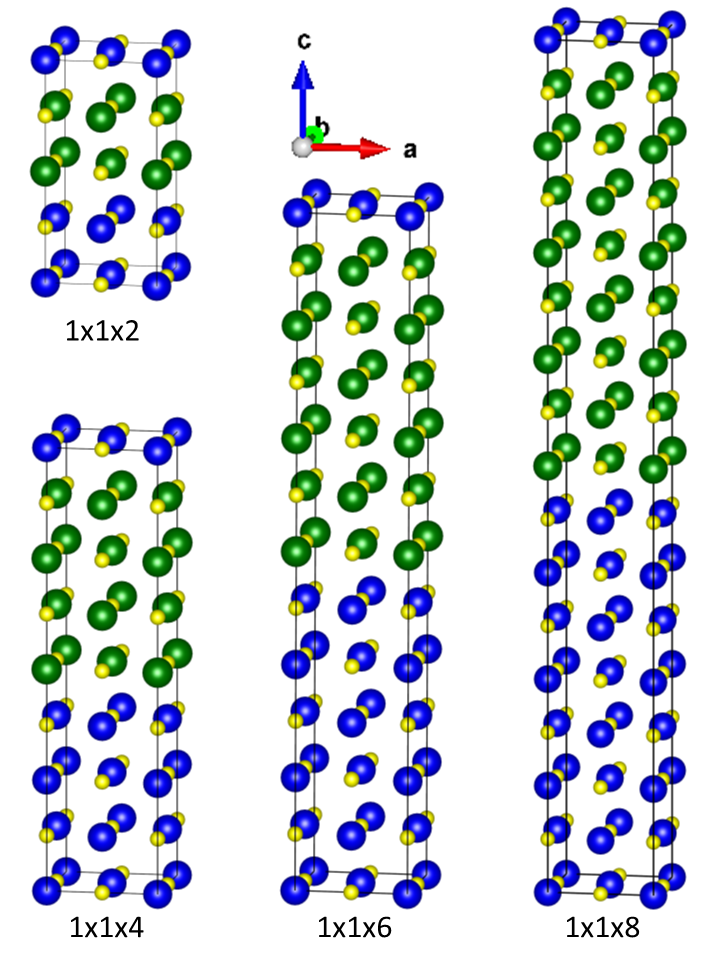}
	\caption{Computational models for $1\times1\times2$, $1\times1\times4$, $1\times1\times6$, and $1\times1\times8$ NaCl-type MoN/TaN superlattices. The yellow, green, and blue spheres correspond to N, Ta, and Mo atoms, respectively. Visualized using the VESTA package\cite{Momma2006,Momma2008,Momma2011}.}
\label{FIG: multilayers}
\end{figure}

The Mo$_{1-x}$Ta$_x$N solid solutions were assumed to adopt the cubic structure with NaCl prototype (Fm$\overline{3}$m, $\#225$, B1-type), often referred to as rocksalt (rs), and the hexagonal structures with NiAs (P6$_3$/mmc, $\#194$), WC (P$\overline{6}$m2, $\#187$), and TaN (and P$\overline{6}$2m, $\#189$) prototypes, respectively.
Supercells with 64 lattice sites (32 metal and 32 nitrogen atoms) were used to model the cubic systems, while supercells consisting in total of 32, 54, and 72 lattice sites modeled the NiAs-type, WC-type, and TaN-type hexagonal systems.
The Ta (Mo) atoms were distributed on the metal sublattice of bulk MoN (TaN) in a disordered manner employing the SQS method \cite{Wei1990Electronic}. 
Additionally, ordered rocksalt Mo$_{1-x}$Ta$_x$N structures were constructed from a conventional cubic B1 cell (8 lattice sites) containing one or two metal atoms of a different type, i.e., one or two Mo (Ta) in 8-atom TaN (MoN), and subsequently expanded to $2\times 2\times 2$ supercells. 
Various (partially) ordered compositions were obtained from the fully ordered $2\times 2\times 2$ supercells with 25\% or 50\% of Mo (Ta) atoms by arbitrarily replacing them by metal atoms of the other type, i.e., Ta (Mo), to obtain the desired Mo-to-Ta ratio. 
The $1\times1\times2$, $1\times1\times4$, $1\times1\times6$, and $1\times1\times8$ NaCl-type supercells consisting of an equal amount of both MoN and TaN shown in Fig.~\ref{FIG: multilayers} served as models for MoN/TaN superlattices with $(001)$ interfaces. 
These supercells were also used to model superlattices with various MoN-to-TaN ratios, i.e., (MoN)$_{1-x}$/(TaN)$_x$, by occupying complete atomic planes by either Mo or Ta atoms and keeping the number of interfaces per supercell. 
The applied periodic boundary conditions produced the superlattice geometry with a bi-layer period in the range of 4.5--18\,\AA.

Lattice parameters of the binary bulk MoN and TaN in the NaCl-, NiAs-, WC-, and TaN-type modifications were optimized by fitting the energy versus volume data with the Birch-Murnaghan equation of state \cite{PhysRev.71.809}, while all structure optimisations in the Mo$_{1-x}$Ta$_x$N and (MoN)$_{1-x}$/(TaN)$_x$ systems were carried out by relaxing supercell volumes, shapes, and atomic positions. 

To compare various systems in terms of their chemical stability, energy of formation, $E_f$, was calculated as
\begin{equation}
E_f=\frac{1}{\sum_s n_s}\bigg(E_{\text{tot}}-\sum_s n_s\mu_s\bigg)\ ,
\end{equation} 
where $E_{\mathrm{tot}}$ is the total energy of the supercell, $n_s$ and $\mu_s$  are the number of atoms and the chemical potential, respectively, of a species $s$. 
The reference chemical potentials for Mo, Ta, and N are conventionally set to the total energy per atom of bcc-Mo, $\mu_{\text{Mo}}$, bcc-Ta, $\mu_{\text{Ta}}$, and N$_2$ molecule, $\mu_{\text{N}}$, respectively.
The mixing enthalpy, $H_{\text{mix}}$, was evaluated according to
\begin{equation}
H_{\text{mix}}=E_{f}-(1-x)E_f^{\text{MoN}}-xE_f^{\text{TaN}}\ ,
\label{Eq: H_mix}
\end{equation}
where $E_f^{\text{MoN}}$ and $E_f^{\text{TaN}}$ are the formation energies corresponding to reference boundary states. 
To find out the thermodynamics of the (MoN)$_{1-x}$/(TaN)$_x$ superlattices, the interface energy, $E_{\text{int}}$, was calculated as
\begin{equation}
E_{\text{int}}=\frac{1}{2A}(E_{\text{tot}}-E_{\text{tot}}^{\text{MoN}}-E_{\text{tot}}^{\text{TaN}}),
\label{Eq: E_int}
\end{equation}
where $A$ is the area of the MoN/TaN interface and $E_{\text{tot}}^{\text{MoN}}$ ($E_{\text{tot}}^{\text{TaN}}$) is the total energy of MoN (TaN) equivalent to that used as a building block in MoN/TaN superlattice.

Furthermore, we investigated elasticity of selected Mo$_{1-x}$Ta$_x$N and (MoN)$_{1-x}$/(TaN)$_x$ systems employing the stress-strain method \cite{le2002symmetry, le2001symmetry}. 
Fourth-order elasticity tensor $\mathbb{C}$, sometimes referred to as stiffness, relates stress, $\sigma$, linearly to strain, $\varepsilon$, according to the Hooke's law
\begin{equation}
\sigma=\mathbb{C}\varepsilon.
\end{equation}
We note that instead of using fourth-order tensor $\mathbb{C}$ in a three-dimensional space, it is often convenient to replace it with a $6\times6$ matrix. 
In the following, $\mathbb{C}$ will represent this matrix of elastic constants in the so-called Voigt's notation.
To evaluate elastic constants corresponding to structures with arbitrary symmetry, we adopt the methodology proposed by \citet{moakher2006closest}. 
First, the squared norm of the elasticity matrix $\mathbb{C}$ is defined as 
\begin{equation}
\|\mathbb{C}\|^2:=\langle \mathbb{C},\mathbb{C}\rangle\ .
\label{Eq: Squarde norm of C}
\end{equation}
The scalar product $\langle \mathbb{C},\mathbb{C}\rangle$ can be calculated in various ways depending on how $\mathbb{C}$ is represented. 
Assuming the Euclidean metrics and the case of 2D representation, Eq.~\ref{Eq: Squarde norm of C} takes form
\begin{equation}
\|\mathbb{C}\|^2 = \sum_{i,j=1}^6C_{ij}^2\ .
\end{equation}
Aiming to simplify the general $6\times 6$ matrix $\mathbb{C}=(C_{ij})$ with 21 independent elements, we wish to project in onto a convenient symmetry class and hence, decrease the number of independent elastic constants. 
Thus we search for a matrix $\mathbb{C}_{\text{sym}}$ of a specific symmetry class such that the norm
\begin{equation}
\|\mathbb{C}-\mathbb{C}_{\text{sym}}\|
\label{Eq: dist(C-Csym)}
\end{equation}
is minimal. 
In other words, we minimize the Euclidean distance between the matrix of elastic constants, $\mathbb{C}$, with an arbitrary symmetry and the elasticity matrix $\mathbb{C}_{\text{sym}}$ of some particular symmetry. 
Rigorous derivation of the projectors for all crystal symmetry classes can be found in Refs. \onlinecite{moakher2006closest,browaeys2004decomposition}. 
Furthermore, according to \citet{mouhat2014necessary}, mechanical stability of a system with an arbitrary symmetry is mathematically equivalent with any of the following necessary and sufficient conditions
\begin{itemize}
\item[(i)] the matrix $\mathbb{C}$ is positive definite,
\item[(ii)] all eigenvalues of $\mathbb{C}$ are positive, 
\item[(iii)] all the leading principal minors of $\mathbb{C}$ are positive,
\item[(iv)] any minor of $\mathbb{C}$ is positive.
\end{itemize} 
Finally, we calculate phonon spectra of selected systems using the Phonopy package \cite{togo2008first}.

\section{Results and Discussion}
\subsection{Chemical stability}
\begin{figure}[h!]
	\centering
    \includegraphics[width=7cm]{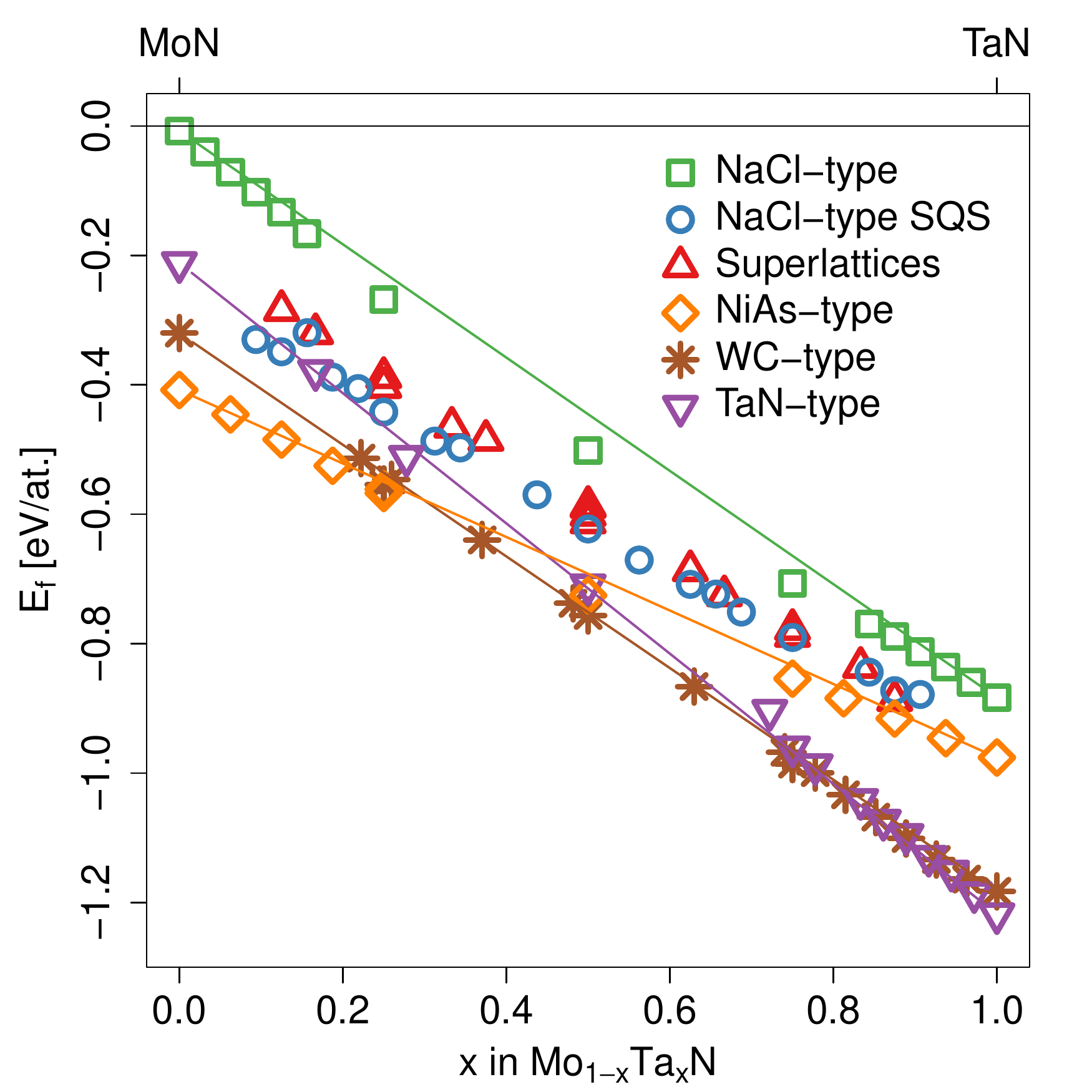}
	\caption{Energy of formation, $E_f$, as a function of $x$ in Mo$_{1-x}$Ta$_x$N solid solutions adopting the cubic NaCl-type, and the hexagonal NiAs-, WC-, and TaN-type structures. 
	Additionally, $E_f$ of the MoN/TaN superlattices is presented. The straight lines connect $E_f$ of the bulk MoN and TaN with the same prototype.
	}
\label{FIG: Ef}
\end{figure}

Figure \ref{FIG: Ef} depicts the energy of formation, $E_f$, as a function of $x$ in cubic (disordered and partially ordered) and hexagonal-structured Mo$_{1-x}$Ta$_x$N solid solutions. 
Clearly, the hexagonal polymorphs are preferred along the MoN--TaN quasi-binary tie-line, i.e., with fully occupied metal and nitrogen sublattices.
In particular, Mo$_{1-x}$Ta$_x$N is predicted to adopt the NiAs-, $\text{WC-,}$ and TaN-type structure for $x$ in intervals of $\sim$(0,0.3), $\sim$(0.3,0.9), and $\sim$(0.9,1), respectively. 
Very close values of $E_f$ obtained for the TaN-type and the WC-type phases for $x$ in the interval of $\sim$(0.9,1) suggest a possible coexistence of these two polymorphs for high Ta content.
When restricting to the cubic-like systems,
the fully disordered Mo$_{1-x}$Ta$_x$N constructed according to the SQS method exhibits the lowest $E_f$, closely followed by the superlattices, in particular for the Ta-rich compositions. 
The partially ordered cubic solid solution is predicted to be the least stable out of the here considered structures.

\begin{figure}[h!]
	\centering
    \includegraphics[width=8cm]{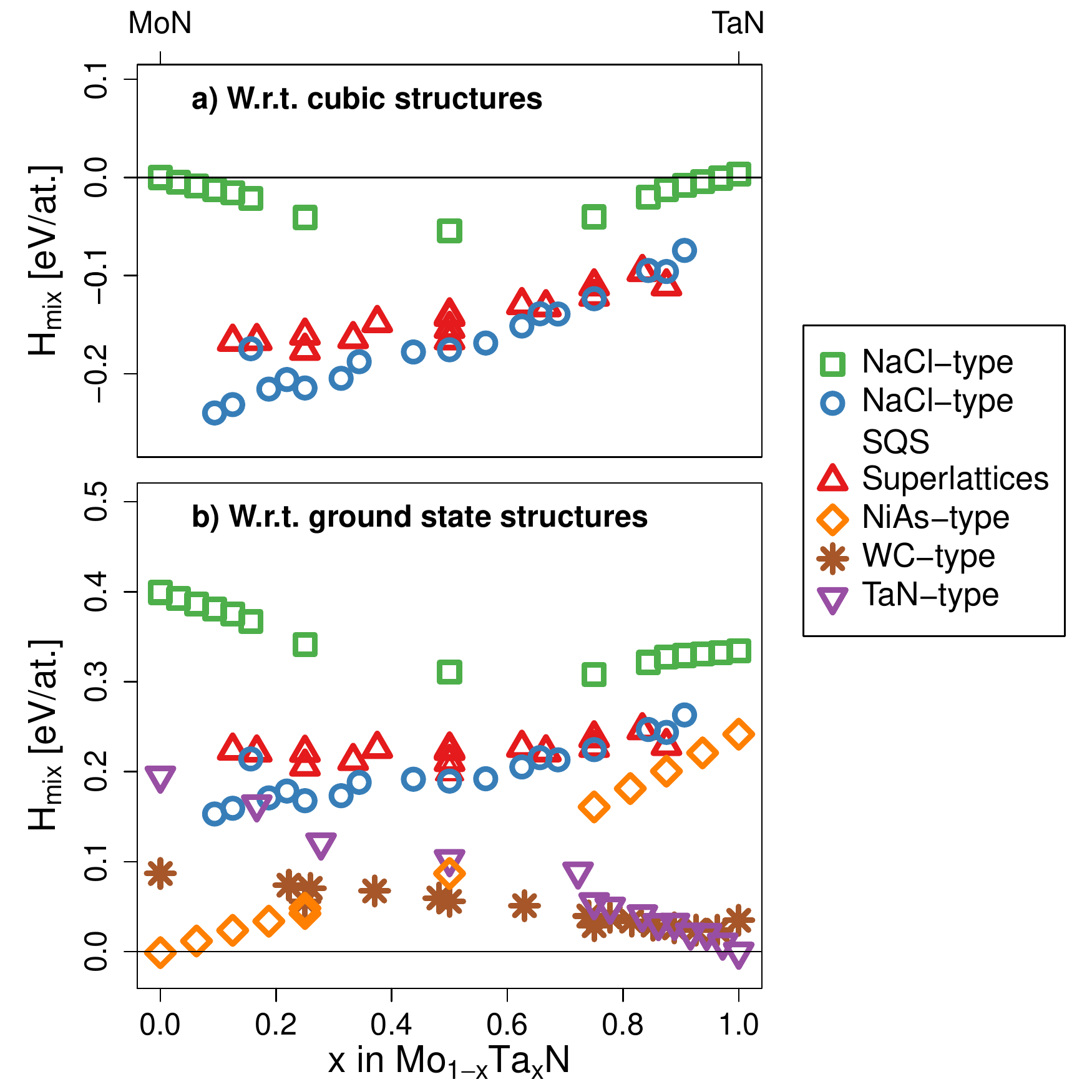}
	\caption{Mixing enthalpy, $H_{\text{mix}}$, as a function of $x$ in Mo$_{1-x}$Ta$_x$N. 
	In panel a), the data points are evaluated with respect to the NaCl-type structures, while in panel b) the zero mixing enthalpy corresponds to ground states of MoN and TaN, i.e., the NiAs-type MoN and the TaN-type TaN.}
\label{FIG: H_mix}
\end{figure}

The mixing enthalpy, $H_{\text{mix}}$, evaluated according to Eq.~\ref{Eq: H_mix} is presented in Fig.~\ref{FIG: H_mix}a for the cubic systems. 
It follows that all cubic solid solutions are predicted to be stable against isostructural decomposition, because their $H_{\text{mix}}$ is negative.
However, since the cubic binary polymorphs are not the respective ground states of MoN and TaN (cf. Fig.~\ref{FIG: Ef}), in Fig.~\ref{FIG: H_mix}b we plot the mixing enthalpy with respect to the NiAs-type MoN and the TaN-type TaN. 
Our results indicate that all Mo$_{1-x}$Ta$_x$N solid solutions together with (MoN)$_{1-x}$/(TaN)$_x$ superlattices are unstable with respect to the ground state hexagonal phases. The calculated $H_{\text{mix}}$ values are in a similar range as for other nitride systems (e.g., Ti$_{1-x}$Al$_x$N or Zr$_{1-x}$Al$_x$N) known to decompose upon thermal loading \cite{Holec2011-la}.

\subsection{Structural analysis of ordered and disordered solid solutions}\label{Subsec: Structural analysis of ordered and disordered solid solutions}
The shape of the $H_{\text{mix}}$ curve for the disordered cubic solid solution is somehow surprising, because it shows an almost linear compositional dependence, which suggests a significant deviation from $H_{\text{mix}}=0$ when extrapolated to $x=0$ (cf.~Fig.\ref{FIG: H_mix}a). 
Since this is supposed to be an isostructural case, $H_{\text{mix}}$ should be $0$ for $x=0$, as Eq.~\ref{Eq: H_mix} implies.
A possible explanation for this deviation is, that due to atomic relaxations present in the solid solution, the structure is significantly modified and hence does not present an isostructural case with the NaCl-type binary boundaries.
Therefore, we analysed local environments of the ordered and disordered solid solutions and compared them with the cubic NaCl-type and the hexagonal $\text{NiAs-,}$ WC-, and TaN-type phases of MoN and TaN. 
A common attribute of all these polymorphs is that every metal atom has 6 nearest N neighbours. 
Hence, 6 distances, $d_1,d_2\ldots,d_6$, and 15 angles, $\varphi_1,\varphi_2,\ldots,\varphi_{15}$ between every (central) metal atom and its nearest N neighbours can be determined, and ordered ascendingly, i.e., in such way that $d_1\leq d_2\leq \ldots\leq d_6$ and $\varphi_1\leq\varphi_2\leq\ldots\leq\varphi_{15}$.  

Let us introduce the $i$th lowest averaged distance, $D_i$,
\begin{equation}
D_i := \frac{1}{M} \sum_{j=1}^M d_i^{\text{at}j}=\text{mean}(d_i^{\text{at1}},d_i^{\text{at2}},\ldots,d_i^{\text{at}M}),
\label{Eq: D_i}
\end{equation}
where $d_i^{\text{at}j}$ denotes the $i$th lowest distance corresponding to the $j$th metal atom in the supercell, i.e., the distance between the $j$th metal atom and its $i$th nearest nitrogen neighbour. 
$M$ is the number of metal atoms included in the analysis.
Similarly, the $k$th lowest averaged angle, $\phi_k$, takes form
\begin{equation}
\phi_k := \frac{1}{M} \sum_{j=1}^M \varphi_k^{\text{at}j}=\text{mean}(\varphi_k^{\text{at1}},\varphi_k^{\text{at2}},\ldots,\varphi_k^{\text{at}M}),
\label{Eq_ varphi_k}
\end{equation}
where $\phi_k^{\text{at}j}$ denotes the $k$th lowest angle corresponding to the $j$th metal atom in the supercell.
Aiming on quantifying the structural (dis)similarity of the disordered cubic Mo$_{1-x}$Ta$_x$N solid solutions with binary prototypes, we further define
\begin{align}
\Delta\text{Dist}:&= \frac{1}{6} \sum_{i=1}^6 \frac{|D_{i}-D_{i}^{\text{Ref}}|}{D_{i}^{\text{Ref}}}= \nonumber \\
&=\text{mean}\bigg(\frac{|D_{1}-D_{1}^{\text{Ref}}|}{D_{1}^{\text{Ref}}}, \ldots,\frac{|D_{6}-D_{6}^{\text{Ref}}|}{D_{6}^{\text{Ref}}}  \bigg),
\label{Eq:Delta Dist}
\end{align}
where $D_{i}$ and $D_{i}^{\text{Ref}}$ correspond to the $i$th lowest averaged distance in the cubic solid solution and in a reference structure, respectively, while the summation runs over all 6 averaged distances. Analogically, we introduce
\begin{align}
\Delta \text{Angles}:&=\frac{1}{15} \sum_{k=1}^{15}\frac{|\phi_{k}-\phi_{k}^{\mathrm{Ref}}|}{\phi_{k}^{\mathrm{Ref}}} \nonumber \\
&=\text{mean}\bigg(\frac{|\phi_{1}-\phi_{1}^{\text{Ref}}|}{\phi_{1}^{\text{Ref}}}, \ldots,\frac{|\phi_{15}-\phi_{15}^{\text{Ref}}|}{\phi_{15}^{\text{Ref}}}  \bigg),
\label{Eq: Delta Angles}
\end{align}
where $\phi_{k}$ and $\phi_{k}^{\text{Ref}}$ denote the $k$th lowest averaged angle in the cubic solid solution and in a reference system, respectively, and the summation runs over all 15 averaged angles. 
It is important to mention that the local environments of Mo and Ta atoms in Mo$_{1-x}$Ta$_x$N were analysed separately, i.e., $M$ in Eqs.~\ref{Eq: D_i} and \ref{Eq_ varphi_k}, was either the number of Mo or Ta atoms; the later (former) case was compared with local environments in binary MoN (TaN) in a corresponding reference structure.

\begin{figure}[h!]
	\centering
    \includegraphics[width=9cm]{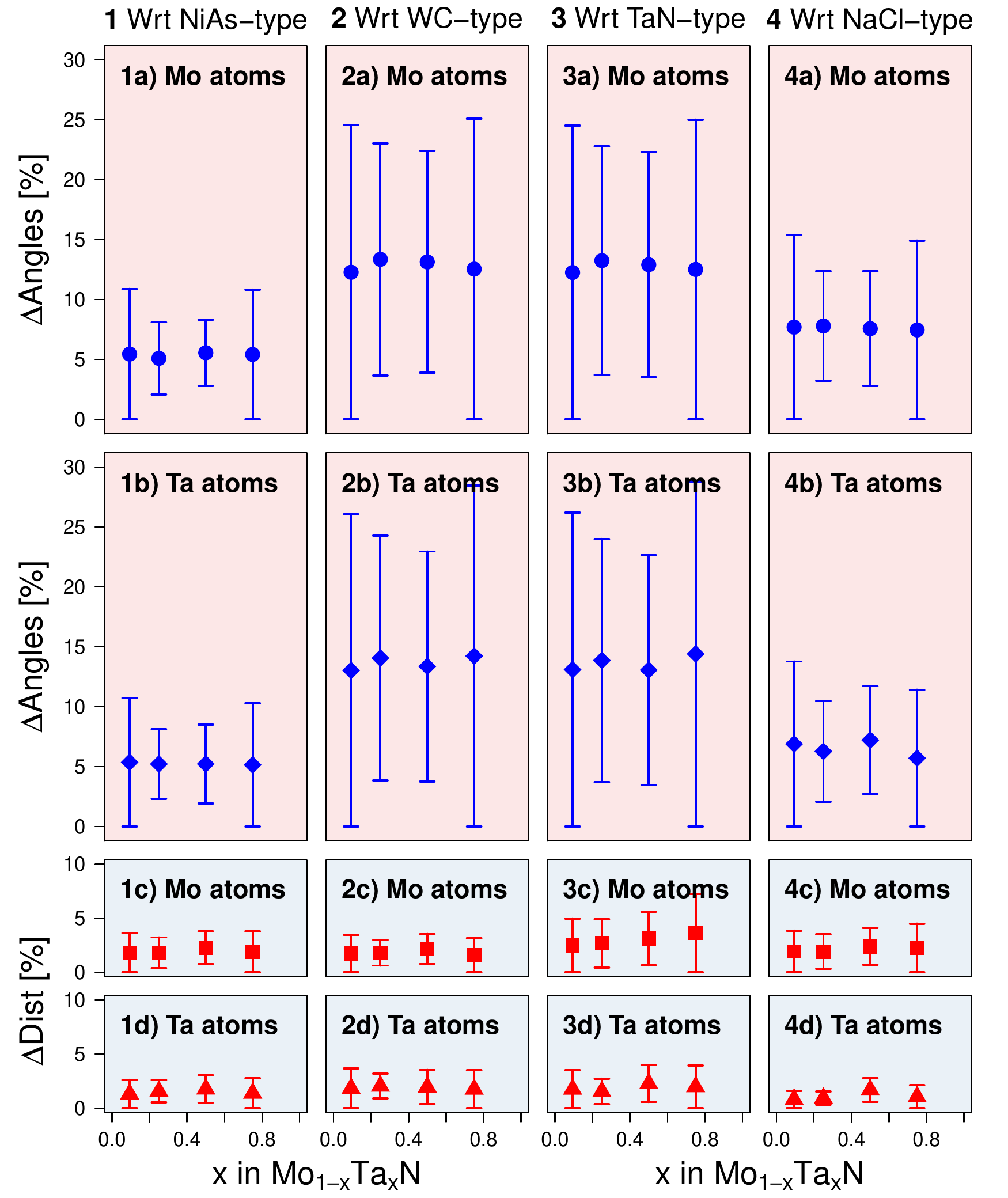}
	\caption{The averaged differences of angles (8 upper panels) and distances (8 lower panels) as a function of $x$ in Mo$_{1-x}$Ta$_x$N. The data evaluated with respect to the hexagonal NiAs-, WC-, TaN-type and the cubic NaCl-type structure are depicted in panels 1a)--1d), 2a)--2d), 3a)--3d), and 4a)--4d), respectively. The error bars are calculated as standard deviations.}
\label{FIG: Local environments}
\end{figure}

Results of the analysis according to the Eqs.~\ref{Eq:Delta Dist} and \ref{Eq: Delta Angles} are plotted in Fig.~\ref{FIG: Local environments}. 
The first interesting observation is that the angles in disordered Mo$_{1-x}$Ta$_x$N differ from all binary systems by about $4$--$15\%$, thus much more significantly than the distances, which differ by $1$--$5\%$. 
Regarding $\Delta\text{Angles}$, the lowest values $\sim 5\%$ are obtained when evaluated with respect to NiAs-type structure, the ground state of MoN. 
Nevertheless, evaluation with respect to cubic NaCl-type polymorph yields very close values as well, with $\Delta\text{Angles}$ between $6$ to $7\%$. 
On the contrary, evaluation with respect to WC-type and TaN-type phases results in significantly high $\Delta\text{Angles}$ of $\sim15\%$, and hence, we conclude that the angles in disordered Mo$_{1-x}$Ta$_x$N are rather dissimilar to these two hexagonal modifications. 
Moreover, the trends are supported also by error bars, which are relatively small for the NiAs- and NaCl-type reference systems, but are almost doubled when referring to the WC- and TaN-type polymorphs. 
We note that the $\varphi_k$ angles are $90\degree$ or $180\degree$ in the cubic rocksalt structure, while they take values of $\sim 82\degree$, $98\degree$, and $180\degree$ in the NiAs-type structure, $\sim 80\degree$, $84\degree$, and $136\degree$ in the TaN-type structure, and $\sim 82\degree$, and $136\degree$ in the WC-type structure. 
Unlike that, $\varphi_k$s in the disordered Mo$_{1-x}$Ta$_x$N are changing gradually from $\sim77\degree$ to $\sim105\degree$, independent of the composition $x$.
Unfortunately, $\Delta {\text{Dist}}$ cannot be interpreted that easily, since all the values are rather small and comparable, regardless of the reference structure.
While $D_i$s evaluated in any of the binary reference structures yield the same value for each system due to the symmetries, the disordered Mo$_{1-x}$Ta$_x$N is characterised by six distinct $D_i$s stemming from the local structural relaxations. 
Nevertheless, the best agreement (by a small margin regarding both the $\Delta {\text{Dist}}$ values as well as the error-bars) is obtained again for the cubic NaCl-type and the hexagonal NiAs-type phase, especially for the Ta atoms.

Therefore, our structural analysis of the disordered cubic Mo$_{1-x}$Ta$_x$N solid solution shows that the symmetries of a rocksalt lattice are partially broken due to the chemical disorder. 
Consequently, the disordered phase is structurally between the cubic NaCl-type and the hexagonal NiAs-type modifications. 
This rationalises the peculiar compositional dependence of $E_f$ corresponding to the disordered cubic Mo$_{1-x}$Ta$_x$N (Fig.~\ref{FIG: Ef}), which is almost perfectly between the $E_f$ values of the (partially) ordered cubic and the NiAs-type hexagonal systems.

\subsection{Superlattice architecture}\label{Subs: superlattice architecture}
\begin{figure}[h!]
	\centering
    \includegraphics[width=5cm]{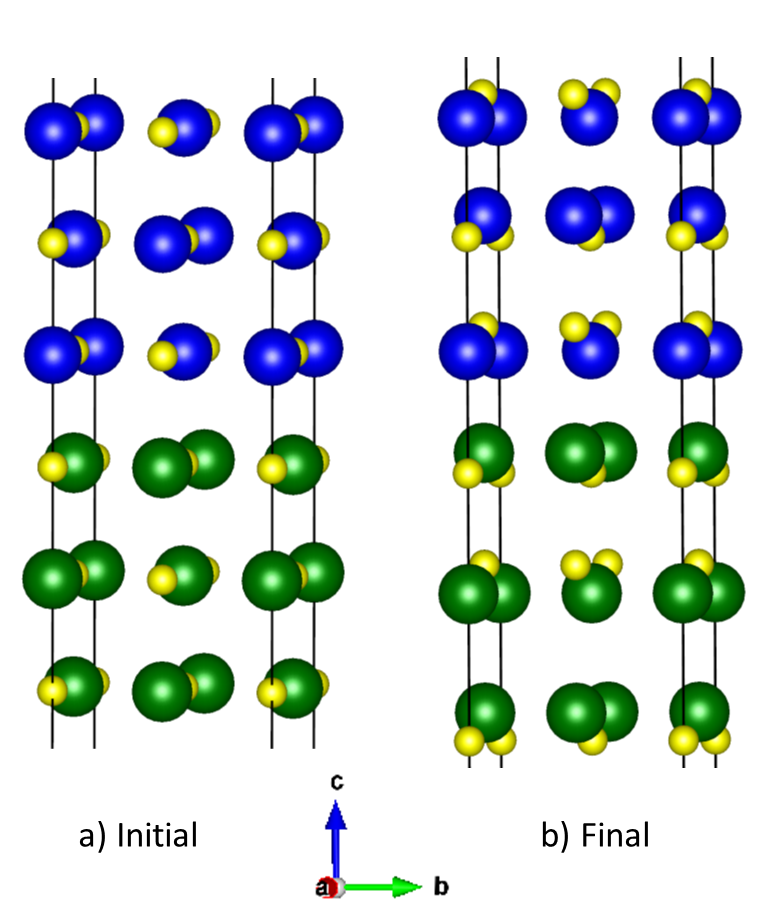}
	\caption{Snapshot of (a) the initial and (b) corresponding fully relaxed structure of a MoN/TaN superlattice. 
	The yellow, green, and blue spheres correspond to N, Ta, and Mo atoms, respectively. Visualized using the VESTA package\cite{Momma2006,Momma2008,Momma2011}.
	}
\label{FIG: Tetragonal deformation}
\end{figure}

The energy of formation of the (MoN)$_{1-x}$/(TaN)$_x$ superlattices was predicted to lie close to $E_f$ of the disordered Mo$_{1-x}$Ta$_x$N solid solutions (Fig.~\ref{FIG: Ef}). 
Furthermore, the corresponding interface energies, $E_{\text{int}}$, are negative with respect to the cubic phases.
These hints indicate that again the symmetries are (partially) broken, i.e., the superlattices are not composed of simply tetragonally-deformed cubic cells. 
Indeed, a further analysis of fully relaxed (MoN)$_{1-x}$/(TaN)$_x$ reveals not only a tetragonal deformation (Fig.~\ref{FIG: Tetragonal deformation}), but also additional atomic displacements, which further break the tetragonal symmetry. 

The tetragonal-like structures of the two binaries, MoN and TaN, were constructed out of $1\times1\times2$ MoN/TaN superlattice, and were fully relaxed until the forces on individual atoms were less than 0.01\,eV/\AA. 
A subsequent structural analysis showed that both these newly proposed phases of MoN and TaN, which will be hereafter referred to as $\zeta\text{-MoN}$ and $\zeta\text{-TaN}$, possess the space group P4/nmm ($\#129$). 
The corresponding lattice parameters together with the atomic positions are summarized in Table~\ref{Tab: Tetragonal phases}. 
The calculated formation energies of $\zeta\text{-MoN}$ and $\zeta$-TaN are $-0.178\,\mathrm{eV/at.}$ and $-0.982\,\mathrm{eV/at.}$, respectively.
Therefore, the $\zeta$-phase is predicted to be more stable than the rocksalt structure for both MoN and TaN, exhibiting $E_f$ of $-0.008\,\mathrm{eV/at.}$ and $-0.887\,\mathrm{eV/at.}$, respectively.

\begin{table*}[h!t!]
\caption{Overview of lattice parameters, $a$ and $c$, and atomic coordinates ($x,y,z$) of the newly identified $\zeta\text{-MoN}$ and $\zeta\text{-TaN}$ phases.}
\centering
\begin{tabular}{ccc|ccc}
\hline
\hline
\multicolumn{3}{c}{$\zeta$-MoN} &
\multicolumn{3}{c}{$\zeta$-TaN} \tabularnewline
 $a, c$ [\AA]  & Atom & ($x, y, z$) &  $a, c$ [\AA]  & Atom & ($x, y, z$)  \tabularnewline
\hline
$a=4.2480$  & Mo1 & $(0.0000, 0.0000, 0.0150)$ &  $a=4.2017$ & Ta1 & $(0.0000, 0.0000, 0.9975)$ \tabularnewline
$c=4.5435$  & Mo2 & $(0.0000, 0.0000, 0.4353)$ &  $c=5.1188$ & Ta2  & $(0.0000, 0.5000, 0.4525)$ \tabularnewline
& Mo3 & $(0.5000, 0.0000, 0.4353)$ &   & Ta3 & $(0.5000, 0.0000, 0.4525)$ \tabularnewline
& Mo4 & $(0.5000, 0.5000, 0.0150)$ &  & Ta4  & $(0.5000, 0.5000, 0.9975)$  \tabularnewline
& N1  & $(0.5000, 0.5000, 0.5311)$ &   & N1  & $(0.5000, 0.5000, 0.5592)$ \tabularnewline
& N2  & $(0.5000, 0.0000, 0.9186)$ &   & N2  & $(0.5000, 0.0000, 0.8907)$  \tabularnewline
& N3  & $(0.0000, 0.5000, 0.9186)$ &   & N3  & $(0.0000, 0.5000, 0.8907)$ \tabularnewline
& N4  & $(0.0000, 0.0000, 0.5311)$ &   & N4  & $(0.0000, 0.0000, 0.5592)$  \tabularnewline   		    		    		    
\hline
\hline
\end{tabular}
\label{Tab: Tetragonal phases}
\end{table*}

\begin{figure}[h!]
	\centering
	\includegraphics[width=8cm]{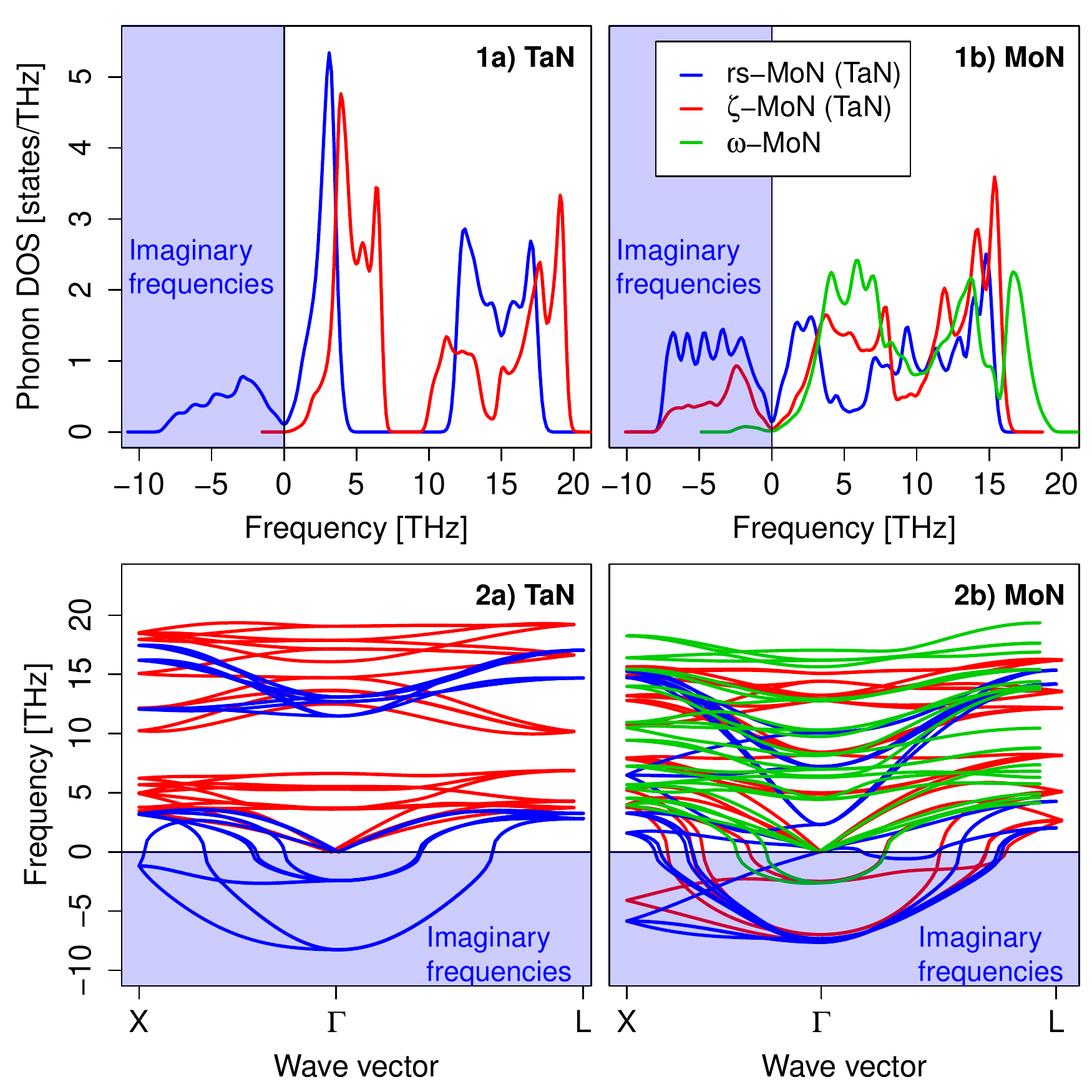}
	\caption{Phonon density of states, panels 1a)--b), and phonon dispersions, panels 2a)--2b), for MoN and TaN adopting the rocksalt (blue lines), tetragonal $\zeta$-type (red lines), and monoclinic $\omega$-type (green lines) structures.}
	\label{FIG: tetragonal structure}
\end{figure}

In order to reveal the driving force for the tetragonal distortion leading to the cubic-to-$\zeta$ transition in MoN and TaN, we calculated their phonon spectra (Fig.~\ref{FIG: tetragonal structure}). 
The phonon density of states and phonon dispersion relation of $\zeta$-TaN do not show any imaginary phonon frequencies (unlike in the case of rs-TaN), hence implies also its mechanical stability (which was further confirmed by testing that the corresponding matrix of elastic constants is positive definite, see Section \ref{Sec:elasticity}). 
Further relaxation following the soft phonon modes in $\zeta$-MoN leads to another low-symmetry structure, a monoclinic $\omega$-MoN (P2$_1$/m, $\#11$) with $E_f$ of about $-0.2454\,\text{eV/at}$, therefore, comparable to that of  $\text{rs-Mo}_{0.91}\text{N}$ \cite{koutna2016point}. 
However, this phase is not relevant for the superlattices as it is forbidden by the superlattice tetragonal symmetry.  

The interface energy, $E_{\text{int}}$ of the (MoN)$_{1-x}$/(TaN)$_x$ superlattices, evaluated with respect to our newly proposed $\zeta\text{-TaN}$ and $\zeta\text{-MoN}$ is close to $0\,\text{eV/\AA}^2$. Moreover, $E_{\text{int}}$ becomes positive, when evaluated with respect to $\zeta\text{-TaN}$ and $\omega\text{-MoN}$. Such values of $E_{\text{int}}$ can be better physically interpreted and agree well with the previously discussed chemical stability (cf. Fig.~\ref{FIG: Ef}).

\subsection{Elastic properties}\label{Sec:elasticity}
Tensors of elastic properties represented using symmetrical $6\times6$ matrices $\mathbb{C}$ were calculated for selected Mo$_{1-x}$Ta$_x$N systems at 5 different compositions, namely, for $x\in\{0,0.25,0.5,0.75,1\}$. 
As these systems belong to different crystal symmetry classes, and moreover, their symmetries are often broken due to the chemical disorder and/or architecture, it is desirable to pay special attention to unifying the method for analyzing their elastic response. 
Therefore, the mechanical stability was tested employing the condition (iii) in Section \ref{Sec: Calculation details}, i.e., that the lowest eigenvalue, $\lambda_{\min}$, corresponding to the (unprojected) matrix $\mathbb{C}$, must be positive.

\begin{figure}[h!]
	\centering
    \includegraphics[width=9cm]{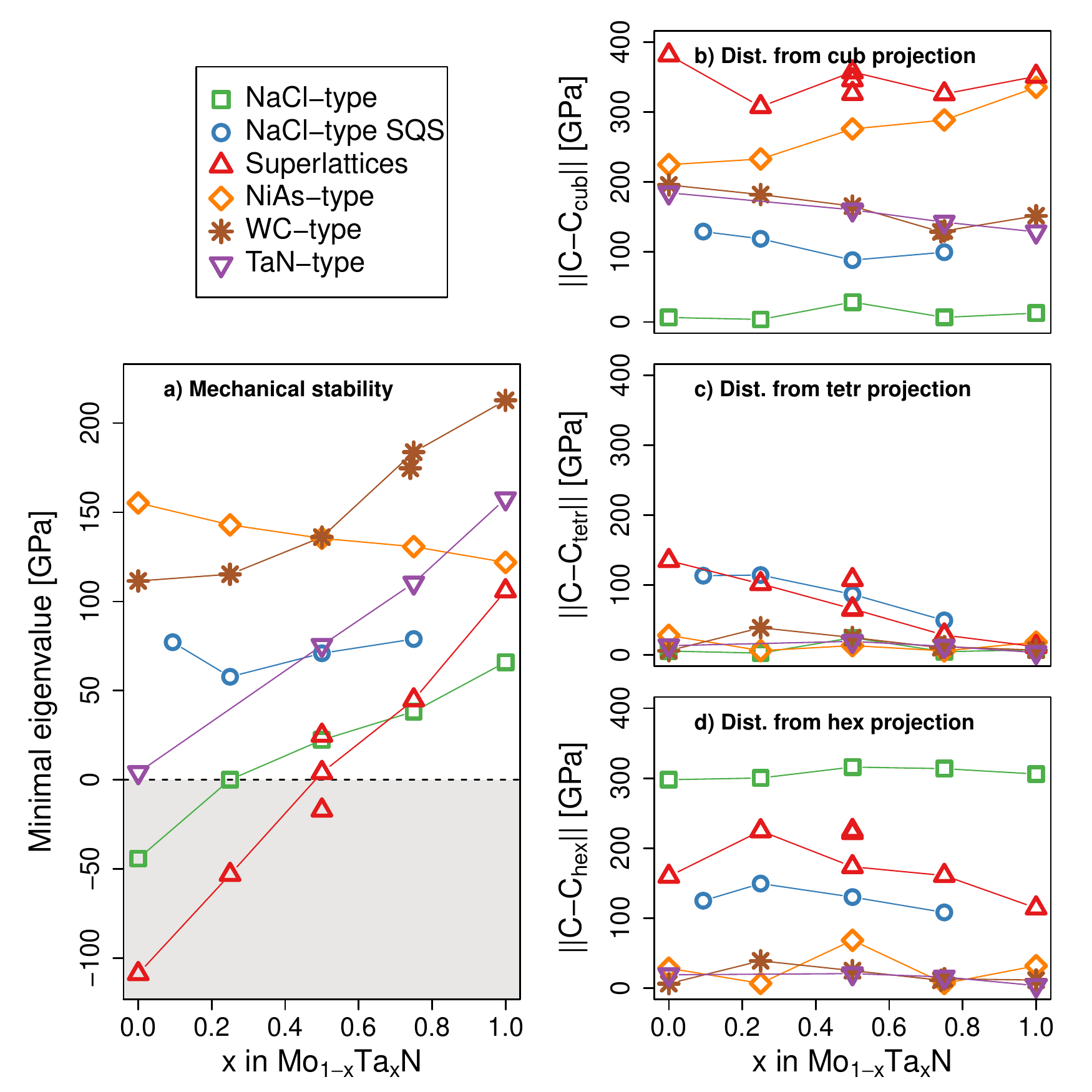}
	\caption{Properties of the elastic tensor corresponding to the Mo$_{1-x}$Ta$_x$N solid solutions and (MoN)$_{1-x}$/(TaN)$_x$ superlattices: minimal eigenvalue (a) together with the distance from cubic (b), tetragonal (c), and hexagonal projection (d).}
	\label{FIG: norms}
\end{figure}

The results presented in Fig.~\ref{FIG: norms}a reveal that all hexagonal Mo$_{1-x}$Ta$_x$N phases are mechanically stable in the whole compositional range, i.e., the corresponding $\lambda_{\min}$ is always positive.
Mechanical stability of the ordered cubic variant depends strongly on the Ta content. 
This finding is consistent with the previous {\it{ab initio}} study by \citet{bouamama2015first} showing that ordered cubic Mo$_{1-x}$Ta$_x$N becomes stable for $x>0.27$. 
On the contrary, the disordered Mo$_{1-x}$Ta$_x$N is predicted to be stabilized by introducing already $\sim 10\%$ of Ta, i.e., for $x\geq 0.1$, which again highlights a strong stabilizing role of the chemical disorder in rocksalt structure.  
Comparable phenomena of stabilizing effects of chemical disorder have been recently shown also for TiAl intermetallic alloys \cite{Holec2015-gs}.
Besides, trends similar to the ordered cubic Mo$_{1-x}$Ta$_x$N were predicted in the case of (MoN)$_{1-x}$/(TaN)$_x$ superlattices, which yield $\lambda_{\text{min}}>0$ for TaN fractions $\geq 50\%$. 
Nevertheless, their mechanical stability strongly depends on the actual bi-layer period: considering a TaN-to-MoN ratio of 1:1, the $1\times1\times2$ superlattice is unstable, the $1\times1\times4$ superlattice is stable, while the $1\times1\times6$ superlattice is again close to instability.
It should be pointed out that all here-considered modifications of TaN, including the newly proposed tetragonal $\zeta\text{-TaN}$, are mechanically stable. 
Unlike that, MoN is predicted to be stable in the hexagonal NiAs-type and WC-type structures, but unstable in the rocksalt and the tetragonal $\zeta$-type modifications, and nearly unstable ($\lambda_{\min}\sim0$) in the hexagonal TaN-type structure (the ground state of TaN). 

In order to decrease the number of generally 21 independent elastic constants (being a consequence of the low symmetry originating from the local chemical disorder), we search for the closest tensor $\mathbb{C}_{\text{sym}}$ with a higher symmetry.
This is done by minimizing the Euclidean distance $\|\mathbb{C}-\mathbb{C}_{\text{sym}}\|$ (cf. Eq.~\ref{Eq: dist(C-Csym)}), thereby representing the best projection of $\mathbb{C}$.
In total we considered three symmetries: cubic, tetragonal, and hexagonal, while the corresponding projections were denoted by $\mathbb{C}_{\text{cub}}$, $\mathbb{C}_{\text{tetr}}$, and $\mathbb{C}_{\text{hex}}$, respectively (Fig.~\ref{FIG: norms}, b--d).
It follows that the ordered Mo$_{1-x}$Ta$_x$N is either cubic, or tetragonal.
The similarity to tetragonal structure may appear surprising at the first sight, but can be easily related to the structural order.
Moreover, $\mathbb{C}_{\text{tetr}}$ has 3 additional degrees of freedom in comparison with $\mathbb{C}_{\text{cub}}$, i.e., $C_{23}\neq C_{12}$, $C_{33}\neq C_{11}$, and $C_{66}\neq C_{44}$, hence, tetragonal projection can minimize distance from $\mathbb{C}$ of cubic system, but the cubic projection will never minimize the distance from $\mathbb{C}$ of tetragonal system.
As a consequence of further reducing the crystal symmetry, e.g., to orthorhombic ($\mathbb{C}_{\text{orth}}$), monoclinic ($\mathbb{C}_{\text{mon}}$) or triclinic ($\mathbb{C}_{\text{tri}}$), the corresponding Euclidean distance can get even smaller.
Regarding the disordered Mo$_{1-x}$Ta$_x$N, the norm $\|\mathbb{C}-\mathbb{C}_{\text{sym}}\|$ does not exhibit any pronounced minimum for either of the high symmetry classes, i.e., $\mathbb{C}_{\text{cub}}$, $\mathbb{C}_{\text{hex}}$, and $\mathbb{C}_{\text{tetr}}$, as it generally yields ``distance'' 50--150\,GPa regardless of the Ta concentration. 
Interestingly, the distances from $\mathbb{C}_{\text{orth}}$, $\mathbb{C}_{\text{mon}}$, and $\mathbb{C}_{\text{tri}}$ of about 40--100\,GPa are also non-negligible.
This can be interpreted as a sign of (partially) broken symmetry, thus supporting our previous conclusions. 
In the case of hexagonal systems, the calculated elastic tensors exhibit large deviations from a strict cubic symmetry, but are very close to hexagonal and tetragonal symmetry. 
The latter is a consequence of the fact that an elastic tensor with the hexagonal symmetry may be viewed as a special case of the tetragonal symmetry \cite{Nye1957-vl}. 
Regarding (MoN)$_{1-x}$/(TaN)$_x$ superlattices, projection onto tetragonal symmetry class is clearly favored over the cubic and hexagonal ones, and this trend becomes more pronounced with increasing Ta concentration, i.e., $\|\mathbb{C}-\mathbb{C}_{\text{tetr}}\|$ decreases from 135\,GPa ($\zeta\text{-MoN}$) to 12\,GPa ($\zeta\text{-TaN}$).

Based on this analysis, the calculated elastic tensors were projected onto the closest tensor of higher symmetry class, excluding trigonal, orthorhombic, monoclinic and triclinic symmetry, and the results are listed in Table \ref{Tab: Projected Cij MoN-TaN}. 
Elastic constants corresponding to the binary MoN and TaN systems are in agreement with the previously published values.

\begin{sidewaystable*}[ht]
    \vspace{8cm}
    \caption{The projected elastic tensors (in GPa) of the Mo$_{1-x}$Ta$_x$N systems and (MoN)$_{1-x}$/(TaN)$_{x}$ superlattices together with $B$, $G$, $E$ moduli (in GPa), Pugh's ratio, $B/G$, and Poisson's ratio, $\nu$. }
    \label{Tab: Projected Cij MoN-TaN}
    \vspace{0.5cm}
    \centering\small\setlength\tabcolsep{2pt}
        \hspace*{-2cm} 
\begin{tabular}{cc|ccccccccccccccccccccc}
\hline
\hline
Phase & $x$ & $c_{11}$ & $c_{11}^{\text{ref}}$ & $c_{12}$ & $c_{12}^{\text{ref}}$ & $c_{13}$& $c_{13}^{\text{ref}}$ & $c_{33}$& $c_{33}^{\text{ref}}$ & $c_{44}$& $c_{44}^{\text{ref}}$ & $c_{16}$  & $B$ & $B^{\text{ref}}$ & $G$ & $G^{\text{ref}}$ & $E$ & $E^{\text{ref}}$ & $B/G$ & $(B/G)^{\text{ref}}$ & $\nu$ & $\nu^{\text{ref}}$ \tabularnewline
\hline
\multirow{7}{*}{NaCl-type}& 0  & 549 & 551$^{\text{\cite{zhao2010displacive}}}$ &212 & 255$^{\text{\cite{zhao2010displacive}}}$ & & & &  & $-43$&$-49^{\text{\cite{zhao2010displacive}}}$&  & -& & -&  &  - && -& &-\\

         &   &  & 543$^{\text{\cite{kanoun2007structure}}}$  & &  171$^{\text{\cite{kanoun2007structure}}}$  & & & &  & & $-73^{\text{\cite{kanoun2007structure}}}$ &  & -& & -&  &  - && -& &-\\

                                         & 0.25 & 612 & & 186 & & & & & & 1  &  &  & 328 && 44 && 125 && 7.52 && 0.44\\
                                         & 0.5  & 651 & & 169 & & & & & & 24 &  &  & 329 && 74 && 205 && 4.48 && 0.40\\
                                         & 0.75 & 687 & & 158 & & & & & & 39 &  &  & 331 && 93 && 256 && 3.55 && 0.37\\
                                         & 1 & 724 & 898$^{\text{\cite{Chang2012Structure}}}$ &147&
                                         131$^{\text{\cite{Chang2012Structure}}}$& & & & & 67 &64$^{\text{\cite{Chang2012Structure}}}$ && 340 & 389$^{\text{\cite{Chang2012Structure}}}$ & 127 & 144$^{\text{\cite{Chang2012Structure}}}$ & 338 & 384$^{\text{\cite{Chang2012Structure}}}$ & 2.69& 2.69$^{\text{\cite{Chang2012Structure}}}$&0.33 & 0.34$^{\text{\cite{Chang2012Structure}}}$\\
 &  & & 678$^{\text{\cite{li2011crystal}}}$ & & 119$^{\text{\cite{li2011crystal}}}$& & & & & & 46$^{\text{\cite{li2011crystal}}}$ && & 306$^{\text{\cite{li2011crystal}}}$ &  & 139$^{\text{\cite{li2011crystal}}}$ &  & 642$^{\text{\cite{li2011crystal}}}$ & & 2.20$^{\text{\cite{li2011crystal}}}$& & 0.15$^{\text{\cite{li2011crystal}}}$\\
\hline
\multirow{4}{*}{NaCl-type SQS}&          0.09 & 469 & & 219 & & & &  & & 99  & &  & 302 && 109 && 291 && 2.78 && 0.34\\
                                       & 0.25 & 513 & & 192 & & & &  & & 94  & &  & 299 && 116 && 209 && 2.57 && 0.33 \\
                                       & 0.5  & 519 & & 205 & & & &  & & 87  & &  & 309 && 110 && 296 && 2.81 && 0.34\\
                                       & 0.75 & 494 & & 227 & & & &  & & 87  & &  & 316 && 103 && 279 && 3.06 && 0.35\\ 
                               
\hline
\multirow{2}{*}{$\zeta$-type}&          0 & 465 & & 144 & & 250 & & 454 & & $-99$ & & 12 &-&&-&&-&&-&&-\\
                                       & 1 & 727 & & 160 & & 154 & & 343 & & 108  & & 0  & 286 && 159 && 403 && 1.79 && 0.26\\                                  
\hline

\multirow{1}{*}{1$\times$1$\times$2 Superlattice}&          0.5 & 689 & & 114 & & 232 & & 440 & & $-9$ & & 1 &-&&-&&-&&-&&-\\

\hline

\multirow{1}{*}{1$\times$1$\times$4 Superlattice}&          0.5 & 696 & & 170 & & 173 & & 407 & & 27 & & 0 &305&&91&&248&&3.35&&0.36\\

\hline
\multirow{3}{*}{1$\times$1$\times$6 Superlattice}&          0.25 & 655 & & 198 & & 172 & & 421 & & $-45$ & & 5 &-&&-&&-&&-&&-\\
                                       & 0.5 & 651 & & 156 & & 191 & & 306 & & 8 & & 0 & 282 && 60 && 168 && 4.70 && 0.40\\                                      
                                       & 0.75& 698 & & 163 & & 173 & & 364 & & 147 & & 0 & 296 && 108 && 289 && 2.74 && 0.34\\                                      
\hline
\multirow{5}{*}{NiAs-type }       & 0  & 491 & 603$^{\text{\cite{zhao2010displacive}}}$ & 180 & 200$^{\text{\cite{zhao2010displacive}}}$& 250 & 205$^{\text{\cite{zhao2010displacive}}}$ & 681& 810$^{\text{\cite{zhao2010displacive}}}$ & 241 &289$^{\text{\cite{zhao2010displacive}}}$  & & 329 & 356$^{\text{\cite{zhao2010displacive}}}$  &188 & 246$^{\text{\cite{zhao2010displacive}}}$  & 474 & 600$^{\text{\cite{zhao2010displacive}}}$ & 1.74 & 1.45$^{\text{\cite{zhao2010displacive}}}$ & 0.26 & 0.22$^{\text{\cite{zhao2010displacive}}}$ \\
                                         & 0.25 & 505 & & 226 & & 220 & & 746 & & 240& &  & 332 && 173 && 442 && 1.91 && 0.28\\
                                         & 0.5  & 519 & & 254 & & 193 & & 764 & & 234 & &  & 340 && 189 && 479 && 1.80 && 0.27\\
                                         & 0.75 & 516 & & 256 & & 176 & & 765 & & 241 & &  & 333 & & 192 && 483 && 1.74 && 0.26\\
                                         & 1    & 531 & 513$^{\text{\cite{zhao2009first}}}$& 268 & 305$^{\text{\cite{zhao2009first}}}$ & 148 & 141$^{\text{\cite{zhao2009first}}}$& 786 & 806$^{\text{\cite{zhao2009first}}}$& 264 & 282$^{\text{\cite{zhao2009first}}}$ &  & 330 & 333$^{\text{\cite{zhao2009first}}}$& 204 & 193$^{\text{\cite{zhao2009first}}}$& 508 & 485$^{\text{\cite{zhao2009first}}}$& 1.61 & 1.73$^{\text{\cite{zhao2009first}}}$& 0.24 & 0.26$^{\text{\cite{zhao2009first}}}$\\
  &  &  & 472$^{\text{\cite{li2011crystal}}}$&  & 270$^{\text{\cite{li2011crystal}}}$ &  & 145$^{\text{\cite{li2011crystal}}}$&  & 726$^{\text{\cite{li2011crystal}}}$& & 243$^{\text{\cite{li2011crystal}}}$ &  &  & 310$^{\text{\cite{li2011crystal}}}$&  & 191$^{\text{\cite{li2011crystal}}}$&  & 669$^{\text{\cite{li2011crystal}}}$& & 1.62$^{\text{\cite{li2011crystal}}}$& & 0.20$^{\text{\cite{li2011crystal}}}$\\
\hline
\multirow{7}{*}{WC-type }& 0    & 600 & 627$^{\text{\cite{zhao2010displacive}}}$ & 179 & 183$^{\text{\cite{zhao2010displacive}}}$ & 221 & 198$^{\text{\cite{zhao2010displacive}}}$& 722 & 788$^{\text{\cite{zhao2010displacive}}}$ & 112 & 122$^{\text{\cite{zhao2010displacive}}}$& & 349 & 353$^{\text{\cite{zhao2010displacive}}}$& 165 & 181$^{\text{\cite{zhao2010displacive}}}$& 428 & 463$^{\text{\cite{zhao2010displacive}}}$& 2.11 & 1.95$^{\text{\cite{zhao2010displacive}}}$& 0.30 &0.28$^{\text{\cite{zhao2010displacive}}}$\\
                                       &   &  & 579$^{\text{\cite{kanoun2007structure}}}$ &  & 147$^{\text{\cite{kanoun2007structure}}}$& &  177$^{\text{\cite{kanoun2007structure}}}$&  & 725$^{\text{\cite{kanoun2007structure}}}$ & &  116$^{\text{\cite{kanoun2007structure}}}$& &  &377$^{\text{\cite{kanoun2007structure}}}$  & & & & 640$^{\text{\cite{kanoun2007structure}}}$& & &&0.24$^{\text{\cite{kanoun2007structure}}}$\\
                                         & 0.25 & 600 & & 175 & & 201 & & 732 & & 129 & & & 341 && 178 && 455 && 1.91 && 0.28 \\
                                         & 0.5  & 606 & & 181 & & 179 & & 752 & & 143 & & & 337 && 189 && 478 && 1.78 && 0.26\\
                                         & 0.75 & 618 & & 188 & & 172 & & 760 & & 182 & & & 339 && 211 && 524 && 1.61 && 0.24\\
                                         & 1    & 635 & 611$^{\text{\cite{Chang2012Structure}}}$  & 203 & 341$^{\text{\cite{Chang2012Structure}}}$ & 147 & 184$^{\text{\cite{Chang2012Structure}}}$ & 804 & 884$^{\text{\cite{Chang2012Structure}}}$ & 232 & 221$^{\text{\cite{Chang2012Structure}}}$ & & 340 & 378$^{\text{\cite{Chang2012Structure}}}$& 239 & 220$^{\text{\cite{Chang2012Structure}}}$& 580 & 554$^{\text{\cite{Chang2012Structure}}}$& 1.42 & 1.72$^{\text{\cite{Chang2012Structure}}}$& 0.22 & 0.26$^{\text{\cite{Chang2012Structure}}}$\\
                                             &    &  & 616$^{\text{\cite{zhao2009first}}}$  &  & 212$^{\text{\cite{zhao2009first}}}$ & & 142$^{\text{\cite{zhao2009first}}}$ &  & 818$^{\text{\cite{zhao2009first}}}$ &  & 256$^{\text{\cite{zhao2009first}}}$ & &  & 337$^{\text{\cite{zhao2009first}}}$&  & 243$^{\text{\cite{zhao2009first}}}$&  & 588$^{\text{\cite{zhao2009first}}}$&  & 1.39$^{\text{\cite{zhao2009first}}}$&  & 0.21$^{\text{\cite{zhao2009first}}}$\\
                                                                                          &    &  & 556$^{\text{\cite{li2011crystal}}}$  &  & 205$^{\text{\cite{li2011crystal}}}$ & & 152$^{\text{\cite{li2011crystal}}}$ &  & 737$^{\text{\cite{li2011crystal}}}$ &  & 222$^{\text{\cite{li2011crystal}}}$ & &  & 318$^{\text{\cite{li2011crystal}}}$&  & 213$^{\text{\cite{li2011crystal}}}$&  & 676$^{\text{\cite{li2011crystal}}}$&  & 1.49$^{\text{\cite{li2011crystal}}}$&  & 0.20$^{\text{\cite{li2011crystal}}}$\\
\hline
\multirow{3}{*}{TaN-type}& 0    & 341 & & 324 & & 219 & & 494 & & 50 & &  & 300 && 34 && 98 && 8.85 && 0.45\\
                                       & 0.5  & 419 & & 267 & & 189 & & 563 & & 114 & & & 299 && 106 && 286 && 2.81 && 0.34\\
                                       & 0.75 & 466 & & 240 & & 180 & & 598 & & 151 & & & 303 && 142 && 368 && 2.14 && 0.30\\
                                       & 1    & 525 & 538$^{\text{\cite{zhao2009first}}}$ & 210 & 238$^{\text{\cite{zhao2009first}}}$& 163 & 165$^{\text{\cite{zhao2009first}}}$& 655 & 665$^{\text{\cite{zhao2009first}}}$& 186 & 191$^{\text{\cite{zhao2009first}}}$& & 308 & 319$^{\text{\cite{zhao2009first}}}$& 182 & 182$^{\text{\cite{zhao2009first}}}$& 456 & 4581$^{\text{\cite{zhao2009first}}}$ & 1.69 & 1.75$^{\text{\cite{zhao2009first}}}$& 0.25& 0.26$^{\text{\cite{zhao2009first}}}$\\
\hline
\hline
\end{tabular}
        \hspace*{-2cm}
\end{sidewaystable*}

Finally, we calculated the polycrystalline bulk ($B$), shear ($G$), and Young's moduli ($E$) using the projected elastic tensors.
We represent them with the Hill's average\cite{hill1952elastic} of the upper bounds according to the Reuss's approach (subscript ``R'')\cite{reuss1929berechnung} and the lower Voigt's bounds (subscript ``V'') \cite{voigt1928lehrbuch}.
General formulae for R and V estimates of $B$ and $G$ can be found in Ref.~\onlinecite{tasnadi2012ab}. 
Young's modulus was evaluated as 
\begin{equation}
E=\frac{9BG}{3B+G}\ .
\end{equation}
The compositional dependence of $B$, $G$, and $E$ (Fig.~\ref{FIG: B,G,E}) for the ordered cubic Mo$_{1-x}$Ta$_x$N are in good agreement with previous {\it{ab initio}} study by \citet{bouamama2015first} reporting that their bulk modulus slightly increases from 306\,GPa (MoN) to 373\,GPa (TaN).
Our calculations yield an increase in $B$ from 325\;GPa (MoN) to 340\;GPa (TaN), which agrees better with the 347\,GPa obtained for NaCl-type TaN by \citet{zhao2009first}. 
The hexagonal NiAs-type and WC-type structures show slightly higher $B$ values.
Specifically, $B$ of the NiAs-structured MoN is approximately $350\,\text{GPa}$, which is comparable with 370\,GPa reported for the cubic boron nitride (c-BN) \cite{haines2001synthesis}, the second hardest material to date.
The bulk moduli of the disordered cubic-like solid solutions are about 30\,GPa below those of the ordered modifications, but comparable to the hexagonal TaN-type system. 
The disordered cubic-like alloys exhibit almost the same shear moduli, $G$, of $\sim 110\,\text{GPa}$, regardless of their Ta content; all other systems show an increase of $G$ with increasing Ta content. 
The shear moduli increase from 188, 165 and $34\,\text{GPa}$ to 204, 239 and $182\;\text{GPa}$ in the case of NiAs-, WC- and TaN-type phases, respectively. 
Here, especially the ordered cubic phase shows a steep increase in $G$ from 44 to $126\;\text{GPa}$ when increasing the
metal-fraction of Ta from 25 at$\%$ (i.e., Mo$_{0.75}$Ta$_{0.25}$N) to 100 at$\%$ (i.e., TaN), which is in good agreement with \citet{bouamama2015first}.

The Young's moduli, $E$, show comparable changes with the composition as the shear moduli, and increase for the hexagonal NiAs-, WC-, and TaN-type systems from 474, 428 and $98\,\text{GPa}$ to 507, 580 and $456\,\text{GPa}$ in the case of NiAs-, WC-, and TaN-type system, respectively. 
The disordered solid solutions exhibit nearly the same $E$ moduli with $\sim294\,\text{GPa}$ for all compositions. Contrary, the ordered cubic phases show a significant increase in $E$ from 125 to $338\,\text{GPa}$, when increasing the metal-fraction of Ta from 25 at$\%$ (i.e.,Mo$_{0.75}$Ta$_{0.25}$N) to 100 at$\%$ (TaN).   
Although $E_f$ of the hexagonal NiAs-, WC- and TaN structured Mo$_{1-x}$Ta$_x$N polymorphs are similar for $x=0.5$ (cf. Fig.~\ref{FIG: Ef}), only the latter two exhibit the highest (and comparable) $B$, $G$, and $E$ moduli. Unlike that, elastic moduli of the TaN-type can be rather compared to the ordered cubic systems and superlattices, which can be rationalized by the fact that MoN in the TaN-type structure is almost mechanically unstable. 

\begin{figure}[h!]
	\centering
    \includegraphics[width=9cm]{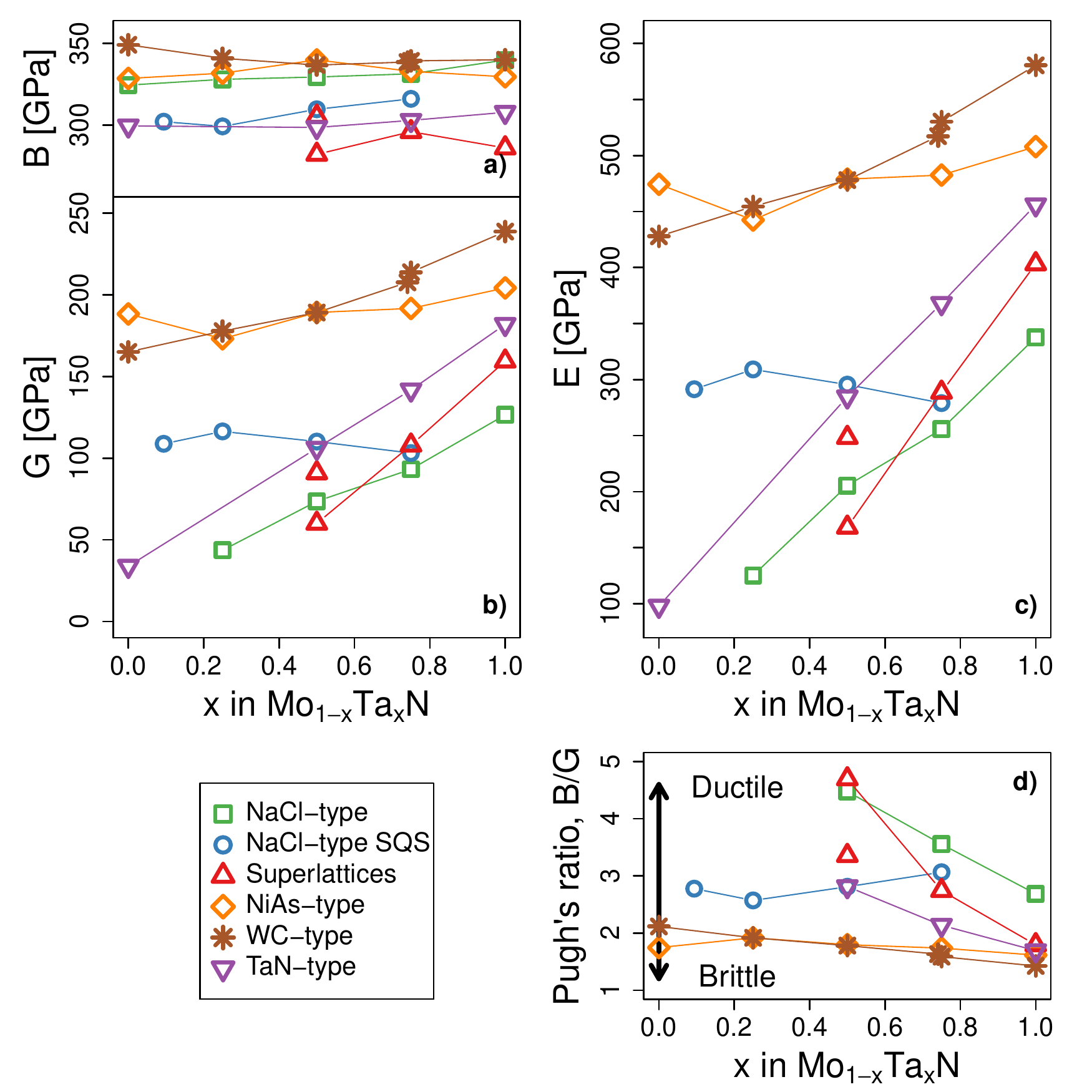}
	\caption{(a) Bulk modulus, (b) shear modulus, (c) Young's modulus of various Mo$_{1-x}$Ta$_x$N solid solutions calculated as Hill's averages of Reuss's and Voigt's polycrystalline isotropic aggregates. 
	(d) Pugh's ratio for estimating relative trends in brittleness/ductility.}
	\label{FIG: B,G,E}
\end{figure}

To rate the ductile/brittle behaviour of our MoN--TaN systems, we used the $B/G$ ratio proposed by Pugh\cite{pugh1954xcii} (also termed as Pugh's ratio). The higher the $B/$G ratio is, the more ductile the material behaves; the critical value separating ductile and brittle materials is $\sim1.75$ \cite{pugh1954xcii}. 
Here, especially the (MoN)$_{1-x}$/(TaN)$_x$ superlattices are highly interesting: although they exhibit the lowest $B$, $G$, and $E$ moduli, they are relatively ductile with $B/G$ of $4.70$ ($1\times1\times6$ superlattice) and $3.35$ ($1\times1\times4$ superlattice) for $x=0.5$, i.e., for equally thick TaN and MoN layers.
Their ductility in terms of $B/G$ is comparable to that of the ordered rocksalt solid solutions.
The disordered cubic alloys exhibit lower $B/G$ ratios for Ta contents above $0.5$, but especially at the Mo-rich side (where the ordered alloys and superlattices are actually mechanically unstable, therefore, we do not show any data points) they provide the highest ductility of all polymorphs considered here.
All three hexagonal modifications yield much lower $B/G$ ratios, and thus are the most brittle structures considered here. 
Finally, we calculated Poisson's ratio, $\nu$,
\begin{equation}
\nu=\frac{3B-2G}{6B+2G},
\end{equation}
and obtained comparable trends in ductile/brittle behavior (cf. Tab.~\ref{Tab: Projected Cij MoN-TaN}) by applying Frantsevich's criterion\cite{frantsevich1983elastic}, which says that the material is brittle for $\nu<1/3$.

\subsection{Electronic density of states} 
\begin{figure}[h!]
	\centering
    \includegraphics[width=8cm]{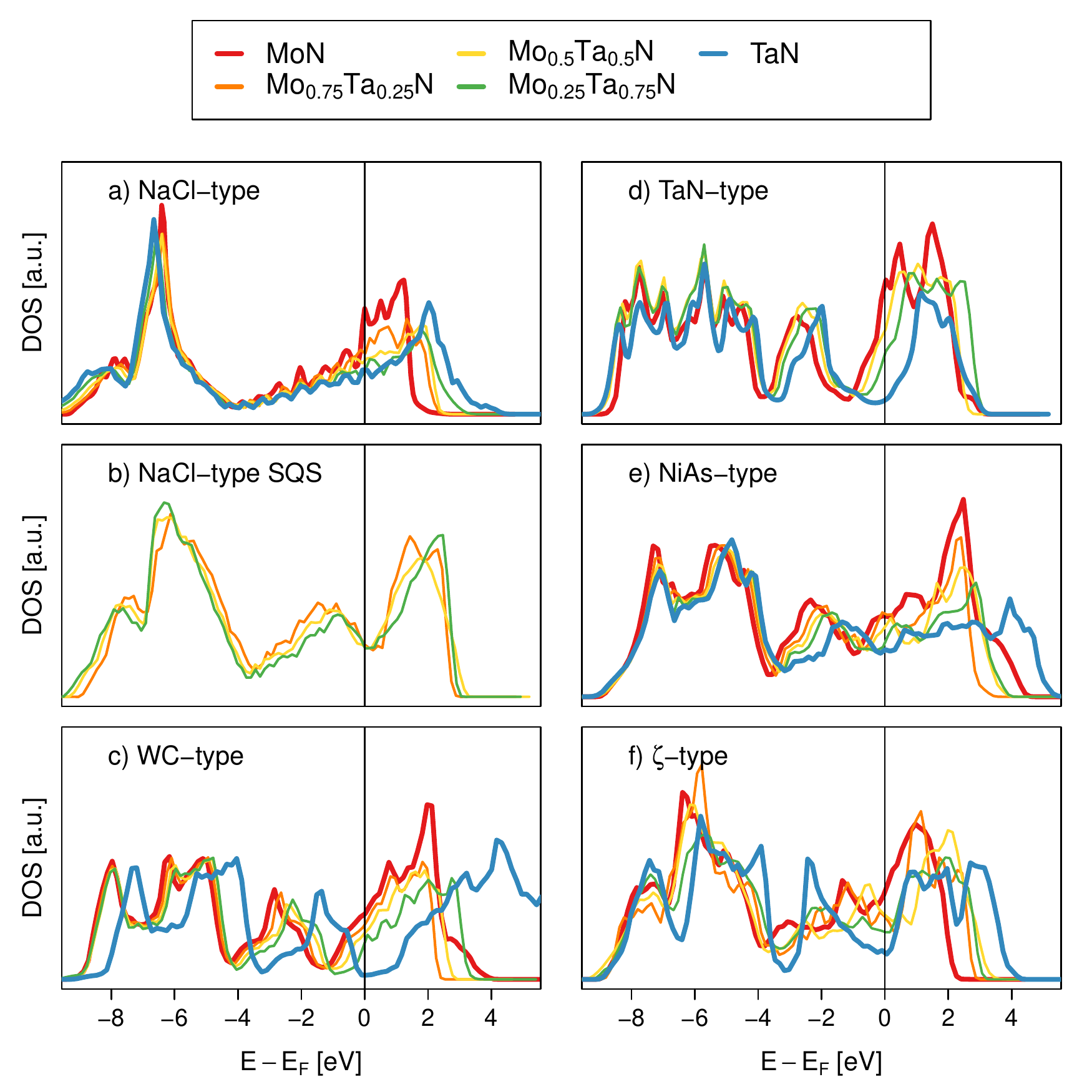}
	\caption{Total density of electronic states, DOS [a.u.], for (a) ordered and (b) disordered NaCl-type, (c) WC-type, (d) TaN-type, (e) NiAs-type, and (d) $\zeta$-type polymorphs of Mo$_{1-x}$Ta$_x$N solid solutions.}
	\label{FIG: DOS}
\end{figure}

Because the total density of electronic states, DOS, is non-zero at the Fermi level, $E_F$, all of our systems are metallic (Fig.~\ref{FIG: DOS}), including the newly identified tetragonal $\zeta$-phases. 
The energy region from $-9$ to about $-4$\,eV can be characterized by strong hybridization of the Mo(Ta)-$d$ electrons with the N-$p$ electrons resulting in a dominant covalent bonding character \cite{Alling2007-bj, Rachbauer2011-lr, Rovere2010-qy}.
The region from $-4\,\text{eV}$ to the Fermi level corresponds to the remaining metal $d$ electrons.
This behavior is well demonstrated by the fact that for every polymorph, the density of states of various compositions are alike; only $E_F$ shifts due to accommodation of the extra electron (per formula unit) in MoN in comparison to TaN.
This band filling leads, however, to pronounced changes around the Fermi level.
All structures of TaN (but the cubic one) exhibit a local DOS minimum at $E_F$ in the case of TaN.
On the contrary, MoN even shows local DOS maximum (peak) for the cubic and hexagonal TaN-type structures.
These trends are perfectly in line with the previously discussed mechanical (in)stability. 
Typically, a distinct peak at the Fermi level is an indicator for structural instability, while a local minimum at $E_F$ suggests that the corresponding system is mechanically stable \cite{Friak-PRB2003,Legut-PRB2010,Zeleny2011,Zhu-2012}.
The disordered alloys show a noticeable increase of DOS around $-1.5$\,eV together with a distinct minimum close to $E_F$ (Fig.~\ref{FIG: DOS}b). This is in strong contrast to the DOS of the ordered solid solutions (Fig.~\ref{FIG: DOS}a).
Therefore, also in terms of their electronic structure, the disordered cubic systems are found dissimilar to the ordered ones. 

\section{Conclusions}
We have carried out extensive first-principles calculations on thermodynamic, structural, mechanical and electronic properties of Mo$_{1-x}$Ta$_x$N solid solutions in their cubic NaCl-type and hexagonal NiAs-, $\text{WC-,}$ and TaN-type structures. 
These are also compared with (MoN)$_{1-x}$/(TaN)$_x$ superlattices with different layer thicknesses and MoN-to-TaN thickness ratios.
The energies of formation clearly indicate that the hexagonal modifications of Mo$_{1-x}$Ta$_x$N are the most stable ones over the whole compositional range, with a preference for the NiAs-, $\text{WC-,}$ and TaN-type structure for composition $x$ in the range of 0--0.3, 0.3--0.9, and 0.9--1.
But especially for $x$ between 0.9 and 1, the WC- and TaN-structures have very similar $E_f$ values.
Consequently, there is no strong preference for either of them and already small changes (like defects) can prefer one over the other. 

Within the metastable cubic structured systems, the disordered solid solutions and the (MoN)$_{1-x}$/(TaN)$_x$ superlattices are significantly more stable than the ordered cubic Mo$_{1-x}$Ta$_x$N solid solutions.
A careful structural analysis of bond lengths and angles clearly suggests that the disordered cubic systems (due to their broken symmetry)  exhibit local similarities with the hexagonal NiAs-type structure.
As soon as Ta is introduced into cubic MoN in a disordered manner, the energy of formation significantly decreases due to these local hexagonal-like environments. Any further addition of Ta only leads to an almost linear change of $E_f$.

Also the (MoN)$_{1-x}$/(TaN)$_x$ superlattices immediately deviate from their cubic symmetry, and the MoN as well as TaN layers relax towards tetragonal structures. Thereby, the formation energy is significantly reduced.
These tetragonal structures, $\zeta$-MoN and $\zeta$-TaN, have the space group P4/nmm ($\#129$) and energies of formation which between their cubic and hexagonal binary polymorphs. 
Importantly, while the cubic phases of MoN and TaN are vibrationally unstable, $\zeta$-TaN is stable and $\zeta$-MoN is less unstable than the cubic MoN.

The disordered cubic systems are also elastically ``between'' the cubic and hexagonal symmetry, due to comparable distances of their elastic tensors from cubic and hexagonal approximants. 
The disordered cubic and all the hexagonal Mo$_{1-x}$Ta$_x$N systems are mechanically stable (although MoN is nearly unstable in TaN-type structure). On the contrary, the ordered cubic systems and the superlattices are mechanically stable only above a critical Ta content of $\sim25\%$ and $\sim50\%$, respectively.    
The polycrystalline elastic moduli suggest that the hexagonal NiAs- and WC-phases of Mo$_{1-x}$Ta$_x$N are significantly harder than the other modifications, but the cubic polymorphs and the sublattices are more ductile (according to Pugh's criterion and Poisson's ratio).
Finally, density of electronic states underpins the conclusions on stability based on energetics and elasticity. 

Our systematic and detailed study on stability, elastic and mechanical properties of various phases along the quasi-binary MoN--TaN system will guide experimental search for functional thin films with outstanding properties.

\section*{Acknowledgements}
This research was supported by the START Program (Y371) of the Austrian Science Fund (FWF) [N.K., D.H., P.H.M.], by the Academy of Sciences of the Czech Republic through the Fellowship of J. E. Purkyn{\v e} [M.F.], the Institutional 
Project No. RVO:68081723, by the Ministry of Education, Youth and Sports of the Czech Republic under the Project CEITEC 2020 (Project No. LQ1601) [M.F., M.{\v S}.], and by the Czech Science Foundation (Project No. GA 16-24711S) [M.F., M.{\v S}.]. 
Computational resources were provided by the Ministry of Education, Youth and Sports of the Czech Republic under the Projects CESNET (Project No. LM2015042), CERIT-Scientific Cloud (Project No. LM2015085), and IT4Innovations National Supercomputer Center (Project No. LM2015070) within the program Projects of Large Research, Development and Innovations Infrastructures [N.K., M.F., M.{\v S}.]. Some calculations were carried out with the help of the Vienna Scientific Cluster 2 (VSC 2) [D.H., N.K.].

\bibliographystyle{apsrev4-1}
\setcitestyle{numbers,square}
\bibliography{biblio}
\end{document}